\documentclass[aps,prd,twocolumn,showpacs,superscriptaddress,groupedaddress]{revtex4}  % for review and submission
\usepackage{graphicx}  % needed for figures
\usepackage{dcolumn}   % needed for some tables
\usepackage{bm}        % for math
\usepackage{amssymb}   % for math
\usepackage{amsmath}   % for math

\begin{document}

% The following information is for internal review, please remove them for submission
%\widetext
%\leftline{Version xx as of \today}
%\leftline{Primary authors: Nguyen Ai Viet}
%\leftline{To be submitted to (PRL, PRD-RC, PRD, PLB; choose one.)}
%\leftline{Comment to {\tt d0-run2eb-nnn@fnal.gov} by xxx, yyy}
%\centerline{\em INTERNAL DOCUMENT -- NOT FOR PUBLIC DISTRIBUTION}

% the following line is for submission, including submission to the arXiv!!
\hspace{5.2in} \mbox{ITI-VNU-Preprint CS-04/2015}

\title{General Relativity in noncommutative spacetime as a unified framework \\ for all interactions and the Higgs field}
\author{Nguyen Ai Viet} \affiliation{ITI, Vietnam National University, Hanoi, Vietnam} \affiliation{Physics Department, College of Natural Sciences, Vietnam National University, Hanoi, Vietnam}
%\author{Nguyen Van Dat}\affiliation{ITI, Vietnam National University, Hanoi, Vietnam}
%\author{Nguyen Suan Han} \affiliation{Physics Department, College of Natural Sciences, Vietnam National University, Hanoi, Vietnam}

%\input authorNAV_list.tex    % D0 authors (remove the first 3 lines
                             % of this file prior to submission, they
                             % contain a time stamp for the authorlist)
                             % (includes institutions and visitors)
\date{\today}

\begin{abstract}
General Relativity in the noncommutative spacetime of ${\cal M}^4 \times Z_2 \times Z_2$ is constructed based on a new type of fermions in addition to the known chiral quark-leptons. While this theory has a finite physical spectrum, all the known interactions of Nature and Higgs fields can be derived from it as components of gravity with only four free parameters.
\end{abstract}

\maketitle

%\section{\label{sec:level1}First-level heading}
% sections are not used for PRL papers

Noncommutative geometry (NCG) proposed by Connes \cite{Connes} is a new concept of spacetime. In particular, in the two-sheeted spacetime ${\cal M}^4 \times Z_2$, Connes and Lott \cite{CoLo} have applied this idea to the Standard Model and derived Higgs fields with a quartic potential, which allows the spontaneous symmetry breaking.

If NCG can be used as the new concept of spacetime in more general context, General Relativity of Einstein must make sense in some particular cases of it. The first attempts to construct the generalized Hilbert-Einstein action in the Connes-Lott's spacetime \cite{nogoNCG} have recovered the Einstein's gravity, but without any new physical content. The spectral action approach \cite{ncgSA} to NCG has provided a unified framework to derive all the interactions encoded in the Dirac operator. However, like Yang-Mills theories, it is just a unified description of interactions.  The real unified theory must base on the first principles of spacetime like the Einstein's theory does. On the other hand, in the spectral action approach, it is not transparent to see how the basic notions of the ordinary Einstein theory have got new meanings in the new spacetime.     

Discretized Kaluza-Klein theory \cite{ncgKK} is an alternative approach to use NCG to unify all the interactions. The use of discrete extra dimensions allows to avoid infinite towers of fields, which exist in other multi-dimensional unified theories. In this theory, the generalized Hilbert-Einstein action of the Connes-Lott's spacetime in this framework contains a finite spectrum of bigravity, bivector and Brans-Dicke biscalar.

Recently, it has been proven \cite{VietDu} that the Hilbert-Einstein action of this theory can contain nonabelian gauge fields only when the gauge group on one sheet must be abelian or two gauge groups must be the same on both sheets. This result could have also been interpreted as an explanation of the  chiral gauge symmetries of QCD and the decoupling of the nonabelian gauge fields from the right-handed quark-leptons. Based on this result, we have proposed to use the noncommutative spacetime of ${\cal M}^4 \times Z_2 \times Z_2$ to unify all known interactions and Higgs fields as components of the generalized gravity \cite{QuyNhon}. The idea is straightforward: The 6D-Hilbert-Einstein action is reduced to the 5D-one together with a 5D $SU(2) \times U(1)$ nonabelian gauge theory. The 5D gauge theory is reduced further to the 4D one of the electroweak coupled to a Higgs doublet with a quartic potential following Connes-Lott's construction. The 5D Hilbert-Einstein action will reduced to the 4D one and the gauge theory of strong interaction.    

%In this Letter, we will investigate this proposal in more details. First, we use a slightly different Dirac operator to obtain a physically meaningful Lagrangian terms, when the coupling to the chiral spinors is considered. The physical content of the generalized Hilbert-Einstein action does not change with this modification. Thus, all the interactions and Higgs fields are unified as components of gravity in this new spacetime, while the coupling to spinor will lead to physically meaningful Lagrangian terms. Secondly, we will choose specific values of Brans-Dicke scalars to derive the physically meaningful results.
Let us construct the generalized Hilbert-Einstein in this spacetime in more details. The noncommutative spacetime ${\cal M}^4 \times Z_2 \times Z_2$ is the usual spacetime extended by two extra discrete dimensions, each having two points. Thus, one must add two more differential elements $DX^5$ and $DX^6$ to the usual 4D ones $dx^\mu$. This spacetime can also viewed as two copies of Connes-Lott's spacetime or four sheets of the ordinary spacetime. While the fifth dimension relates the chiral particles, the physical meaning of the sixth dimension will need more careful analysis to be clarified.

The basic building blocks of NCG in our four-sheeted noncommutative spacetime are the following spectral triplet:

i) The Hilbert space ${\mathcal H}= {\mathcal H}^v \oplus {\cal H}^w $ is a direct sum of two Hilbert spaces ${\mathcal H}^u = {\cal H}^u_L \oplus {\cal H}^u_R; u = v,w$, which in turns are direct sums of the Hilbert spaces of left-handed and right-handed spinors. Thus the wave functions $ \Psi \in {\mathcal H}$  can be represented as follows 
\begin{equation}
    \Psi(x) =  \begin{bmatrix}
    \Psi^v(x) \cr
    \Psi^w(x) 
    \end{bmatrix} ~~,~~ 
    \Psi^u(x) = \begin{bmatrix}
        \psi^v_L(x) \cr
        \psi^w_R(x) 
      \end{bmatrix}  \in {\mathcal H}^u ~;~ u=v,w,
\end{equation}
where the chiral spinors $\psi^u_{L,R}(x) \in {\cal H}^u_{L,R}$ are defined on the 4D spin manifold ${\cal M}^4$. The $v$-type chiral spinors will be identified as the usual quark-leptons, while the $w$-type ones are new matter fields.

ii) The algebra ${\cal A}={\cal A}^v \oplus {\cal A}^w ; {\cal A}^u = {\cal A}^u_L \oplus {\cal A}^u_R$ contains the diagonal $4 \times 4$ matrix 0-forms ${\cal F}$  
\begin{equation}\label{0form}
   {\mathcal  F}(x) =  \begin{bmatrix}
    F^v(x) & 0 \cr
    0 & F^w(x) 
    \end{bmatrix}
     ~,~ F^u(x) =  \begin{bmatrix}
        f^u_L(x) & 0 \cr
        0 & f^u_R(x) 
        \end{bmatrix}
         \in {\cal A}^u,
\end{equation}
where $f^u_{L,R}(x)$ are real valued operators acting on the Hilbert spaces ${\cal H}^u_{L,R}$.

iii) The Dirac operator ${\cal D} = \Gamma^P \partial_P = \Gamma^\mu \partial_\mu + \Gamma^5 \partial_5+ \Gamma^6 \partial_6, P=0,1,2,3,5,6 $ is defined as the following $4 \times 4$ matrix operator
%\begin{widetext}
\begin{eqnarray}\label{Dirac1}
{\cal D} = \begin{bmatrix}
D & im_1 \theta \otimes {\bf e}\cr
-i m_1 \theta \otimes {\bf e}  & D 
\end{bmatrix},~
D = \begin{bmatrix}
\gamma^\mu \partial_\mu & i m_2 \gamma^5 \cr
- i m_2 \gamma^5 & \gamma^\mu \partial_\mu
\end{bmatrix}&& 
\end{eqnarray}
%\end{widetext}
where $\gamma^\mu$ and  $\gamma^5$ are the usual Dirac matrices, $\theta$ is a self-adjoint Clifford element $ \theta^2=1$, $m_1, m_2$ are mass parameters. This Dirac operator will give physically meaningful coupling of gravity to chiral spinors as shown in Ref.\cite{VW2015}.

The partial derivatives $\partial_5$ and $\partial_6$ over the discrete dimension are defined as follows
\begin{eqnarray}
\partial_6 {\cal F} &=& \sigma^\dagger \otimes {\bf e} [D_6, {\cal F}]=  m_1(F^v - F^w){\bf r} \otimes {\bf e} ~,~ \\
\partial_5 F^u & =& \sigma^\dagger [D_5, F^u] = m_2(f^u_L-f^u_R) {\bf r},
\end{eqnarray}
where
\begin{eqnarray}
 D_5 & = & \begin{bmatrix}
 0 & m_1 {\bf e} \cr
 -m_1 {\bf e} & 0
 \end{bmatrix} ~,~ 
D_5 =  \begin{bmatrix}
 0 & m_2 \cr
 -m_2 & 0
 \end{bmatrix} 
 \\
{\bf e} & = & \begin{bmatrix}
1 & 0 \cr
0 & 1
\end{bmatrix} ~,~
{\bf r} = \begin{bmatrix}
1 & 0 \cr
0 & -1 
\end{bmatrix} ~,~ \sigma^\dagger =
\begin{bmatrix}
0  &  -1\cr
1 & 0
\end{bmatrix}.  
\end{eqnarray} 
The exterior derivative of a 0-form ${\cal F}$ is given in the following $4 \times 4$ matrix form
%\begin{widetext}
\begin{eqnarray} \label{DER0FORM}
{\cal D}{\cal F}  = & \begin{bmatrix}
DF^v  & i m_1 \theta(F^v- F^w) \cr
i m_1 \theta (F^v-F^w) & DF^w
\end{bmatrix} & \nonumber \\ 
DF^u  = & \begin{bmatrix}
df^u_L  & i m_2\gamma^5 (f^u_L - f^u_R) \cr
i m_2 \gamma^5 (f^u_L- f^u_R) & df^u_R %& 0 \nonumber \\
%im_2 \gamma^5 & df^v_R  &  0   &  im_1 \theta (f^v_R - f^w_R)   \cr
%i m_1 (f^v_L - f^w_L)& 0 & df^w_L & im_2 \gamma^5(f^w_L - f^w_R) \cr
% 0 & i m_1(f^v_R - f^w_R) & im_2\gamma^5 (f^w_L - f^w_R) & df^w_R
\end{bmatrix} &
\end{eqnarray}
%\end{widetext}
The generalized 1-forms are defined as follows  
\begin{eqnarray}
{\cal U} & = & \begin{bmatrix}
U^v & i\theta U^w_6 \cr
i\theta U^v_6 & U^w
\end{bmatrix},
\end{eqnarray}
where $U^u_6$ are 5D $2 \times 2$ matrix 0-forms and $U^u$ are 5D $2 \times 2$ 1-form defined as follows
\begin{eqnarray}
U^u_6 & = & \begin{bmatrix}
u^u_{6L} & 0 \cr
0  & u^u_{6R}
\end{bmatrix} ~,~ U^u = \begin{bmatrix}
u^u_L   &  i \gamma^5 u^u_{5R} \cr
i \gamma^5 u^u_{5L} & u^u_R
\end{bmatrix},
\end{eqnarray} 
where $u^u_L, u^u_R$ are 4D 1-forms, $u^u_{5L,R},u^u_{6 L,R}$ are 4D 0-forms. 

The module of 1-forms are spanned by the $4 \times 4$ Dirac matrices $\Gamma^P$ as follows
\begin{eqnarray}
{\cal U} = \Gamma^P U_P ~,~
\Gamma^\mu  =
\begin{bmatrix}
\gamma^\mu \otimes {\bf e} & 0 \cr
0 & \gamma^\mu \otimes {\bf e}
\end{bmatrix}, && \nonumber \\
\Gamma^5  = 
\begin{bmatrix}
i \gamma^5 \otimes \sigma & 0 \cr
0  & i \gamma^5 \otimes \sigma
\end{bmatrix} ~,~
\Gamma^6  = \begin{bmatrix}
0 & i\theta \otimes {\bf e} \cr
-i \theta \otimes {\bf e} & 0
\end{bmatrix}, &&
\end{eqnarray}

In order to avoid the "junk forms", we use the anti-commutative wedge product of 1-form defined as follows
\begin{equation}
\Gamma^P \wedge \Gamma^Q = - \Gamma^Q \wedge \Gamma^P = {1 \over 2} [\Gamma^P, \Gamma^Q]
\end{equation}
A generalized 2-form ${\cal S}$ is spanned in this basis as follows
\begin{equation}
{\cal S} = \Gamma^P \wedge \Gamma^Q S_{PQ}~,~ S_{PQ} = -S_{QP} 
\end{equation} 
where $S_{PQ}$ are 0-forms.

The generalized Cartan's structure equations are defined in a perfect parallelism with the ordinary ones.
\begin{eqnarray}
{\cal T}^E &= &  DE^E + E^F \Omega^E_F \label{Torsion}\\
{\cal R}^{EF} &=& D\Omega^{EF}+ \Omega^E_G \wedge \Omega^{GF} \label{Curv},
\end{eqnarray}
where ${\cal T}^E$ and ${\cal R}^{EF}$ are the generalized torsion and curvature 2-forms, ${\cal E}^E$ and $\Omega^{EF}$ are the generalized vielbein and Levi-Civita connection 1-forms. 

For our convenience, the following convention for the 6D, 5D and 4D flat and curved indexes is used
\begin{eqnarray}
E,F,G = A, \dot{6}; A,B,C = a, \dot{5}; a,b,c =0,1,2,3 &&\nonumber \\
P,Q,R = M,6; M,N,L =\mu, 5;\mu,\nu, \rho=0,1,2,3.&&
\end{eqnarray}

The basis elements ${\cal E}^E$ of the locally flat reference frame are linear transformations of the curvilinear one $DX^P$ with the vielbein coefficients $E^P_E(x)$ as follows
\begin{eqnarray}
{\cal E}^E = DX^P {\cal E}^E_P(x) &~,~&  DX^P = {\cal E}^E {\cal E}^P_E(x) \\
{\cal E}^P_E(X) {\cal E}^E_Q(x) = \delta^P_Q &~,~&  {\cal E}^P_E(x) {\cal E}^F_P(X) = \delta^E_F 
\end{eqnarray}

The generalized metric tensor can be introduced via the vielbein coefficients as follows
\begin{eqnarray}
{\cal G}^{PQ}(x) &=& {\cal E}^P_E \eta^{EF} {\cal E}^Q_F ~,~
{\cal G}_{PQ}(x) = {\cal E}_P^E \eta_{EF} {\cal E}_Q^F,
\end{eqnarray}
where $\eta^{EF} = diag(-1,1,1,1,1,1)$. 

The Levi-Civita connection 1-forms $\Omega^\dagger_{EF} = - \Omega_{FE}$ are introduced as a direct generalization of the ordinary ones. With a generalized torsion free condition \cite{ncgKK}, one can determine all the Levi-Civita connection 1-forms and hence the Ricci curvature tensor from the generalized Cartan structure equations in terms of vielbeins.

The Ricci scalar curvature ${\cal R}_6=\eta^{EG} \eta^{FH} {\cal R}_{EFGH}$ and  the generalized Hilbert-Einstein action are direct generalizations of the ordinary notions
\begin{eqnarray}\label{SHE}
{\cal S}_{HE}(6) \sim M^2_{Pl} \int dx^6 \sqrt{-det|{\cal G}|} {\cal R}_6, 
\end{eqnarray}
where $M_{Pl}$ is the Planck mass, the integration over the discrete dimension can be replaced by the trace.

In this paper we will use the following generalizations of $det|g|$ in 5D and 6D
\begin{equation}
det |{\cal G} = Tr_6(det |G| G_{66}) ~,~ det |G| = Tr_5(det|g| G_{55}).
\end{equation}

The dimension reduction is carried out to derive the 4D action from ${\cal S}_{HE}(6)$ in two subsequent steps. In the first step, the 6D objects are reduced to the 5D ones as follows:    

Without any loss of generality, the most general 6D vielbein coefficients are given as follows
\begin{eqnarray}
{\cal E}^A_M &=& \begin{bmatrix}
E^{vA}_M & 0 \cr
0 & E^{wA}_M
\end{bmatrix} ~,~ {\cal E}^A_6 = 0\nonumber \\
 {\cal E}^{\dot 6}_M &=& \begin{bmatrix}
 A^{v}_M & 0 \cr
 0 & A^{w}_M
 \end{bmatrix} ~,~
 E^{\dot{6}}_6 = \begin{bmatrix}
 \Phi^v & 0 \cr
 0 & \Phi^w
 \end{bmatrix},
\end{eqnarray}
where $E^{uA}_M$ are a pair of 5-d vielbeins, $A^u_M$ are a pair of 5-d vectors, $\Phi^u$ are a pair of Brans-Dicke scalars.

With the following choice of the vielbein
\begin{equation}
E^{vA}_M = E^{uA}_M ~,~ \Phi^v = \Phi^u = 1,
\end{equation}
the 6D Ricci scalar curvature is reduced to
\begin{eqnarray}
&{\cal R}_6  = R_5+ {\cal L}_g= R_5 - {1 \over 4} G^{MK} G^{NL} \hat{J}_{MN} \hat{J}^\dagger_{KL},& \label{LG5}\\
&\hat{J}_{MN} = \partial_M A_{+N}- \partial_N A_{+M} 
+ 2m_1 [A_{-M}, A_{-N}]& \label{gstrength},
\end{eqnarray}
where $A_{\pm M} = {1 \over 2}(A^v_M \pm A^w_M)$. The gauge field $ A^w_M $ must be abelian to keep $L_g(5)$ gauge invariant as follows`
\begin{eqnarray} 
L_g(5) &=& - {1 \over 4} G^{MK} G^{NL} (J^\dagger_{MN} J_{KL}+ K^\dagger_{MN} K_{KL})\label{LG5} \\
J_{MN} &=& \partial_M A^v_{N}- \partial_N A^v_M 
+ 2m_1 [A^v_{M}, A^v_{N}]  \nonumber \\
K_{MN} &=& \partial_M A^w_N - \partial_N A^w_M 
\end{eqnarray}
In the second dimension reduction step, $R_5$ and ${\cal L}_g(5)$ are reduced to 4D terms. The gravity sectors of $w$ and $v$ types are assumed to be the same  
\begin{equation}
E^{v A}_M  =  E^{w A}_M = E^A_M 
\end{equation}
Let us follow Ref.\cite{VietDu} to specialize with the following ansatz
\begin{eqnarray}\label{5vielbein}
E^a_\mu & = & e^a_\mu {\bf e} ~,~ E^{a}_5 = E^\mu_{\dot {5}} =  0 ~,~ E^\mu_a = e^\mu_a {\bf e} \nonumber \\
E^{\dot{5}}_\mu & = & A_\mu = {1 \over M}\begin{bmatrix}
\alpha  C^i_{\mu} \lambda^i & 0 \cr
0 & \beta C^i_{\mu} \lambda^i
\end{bmatrix} ~,~ E^{\dot{5}}_5 = \Phi = \begin{bmatrix}
\alpha & 0 \cr
0 & \beta
\end{bmatrix},\nonumber \\
E^5_a &=& {1 \over M}\begin{bmatrix}
e^\mu_a C^i_{\mu} \lambda^i & 0 \cr
0 & e^\mu_a C^i_{\mu} \lambda^i
\end{bmatrix} ~,~ E^5_{\dot{5}} = \begin{bmatrix}
\alpha^{-1} & 0 \cr
0 & \beta^{-1}
\end{bmatrix},
\end{eqnarray}
where $\lambda^i, i=1,..., 8 $ are $3 \times 3$ GellMann matrices, $C^i_\mu(x)$ are $SU(3)$ gauge fields. The mass parameter $M$ and the constants $\alpha, \beta$ will be determined later. 

The metric of this vielbein is determined as follows
\begin{eqnarray}\label{METRIC}
G^{\mu\nu} &=&  g^{\mu \nu} {\bf e} ~,~ G_{\mu\nu} = g_{\mu \nu} {\bf e} + {1 \over M^2}C_\mu C_\nu \Phi^2, \nonumber \\
G^{\mu 5} &=&  {1 \over M}g^{\mu \nu} C_\nu \Phi = G^{5 \mu} , \nonumber \\
G^{55} &=& \Phi^{-2}+ {1 \over M^2}g^{\mu \nu} C_\mu C_\nu \Phi^2, \nonumber \\
G_{\mu 5} &=&  G_{5\mu}= {1 \over M}C_{\mu}\Phi^2~,~ G_{55} = \Phi^2 
\end{eqnarray}

Following Ref.\cite{VietDu}, we can reduce $R_5$ to the following $SU(3)$ gauge invariant form
\begin{eqnarray}
R_5 &= & r_4 - {1 \over 4} {1 \over M^2_{Pl}} H^{\mu \nu} H_{\mu \nu} \label{r5}\\
H_{\mu \nu} &=& \partial_\mu C_{\nu} - \partial_\nu C_{\mu} + g_S [C_{\mu}, C_{\nu}], \\
g_S & = & 2{m_2 \over M} {1- \tan\theta_S \over 1 + \tan^2\theta_S} \\
\alpha & =& {2 M \over M_{Pl}}\cos\theta_S ~,~ \beta = {2 M \over M_{Pl}} \sin\theta_S
\end{eqnarray}

So, the 5D gravity sector of ${\cal S}_{HE}(6)$ is reduced to the following action of QCD coupled to 4D gravity
\begin{eqnarray}
& S_{QCD} &= M^2_{Pl} \int dx^6 \sqrt{-det|{\cal G}|} R_5 \nonumber \\
&=&  \int dx^4 \sqrt{-det|g|} (M^2_{Pl}r_4 - {1 \over 4} H^{\mu \nu} H_{\mu \nu}).  
\end{eqnarray}
The model has three free parameters $m_2, M$ and $\theta_S$. However, as we will see later, $M$ and $\theta_S$ are determined when we couple gravity to the chiral spinors.

Now we can follow Connes-Lott \cite{CoLo} to reduce the 5D gauge Lagrangian ${\cal L}_g(5)$ to a 4D one. For our purpose, we do not consider the $w$-type gauge fields and choose the v-type ones $A^v_M$ as follows
\begin{eqnarray}
A^v_\mu &=& \begin{bmatrix}
A^v_{\mu L} & 0 \cr
0 & A^v_{\mu R} 
\end{bmatrix} ~,~ A^v_5 = \begin{bmatrix}
\varphi & 0 \cr
0 & \varphi^\dagger
\end{bmatrix}.
\end{eqnarray}
The physical electroweak gauge vector fields $E_\mu(x),  W^i_\mu(x), i=1,2,3$ and the Higgs fields $H^\alpha$ can be introduced as follows
\begin{eqnarray}
A^v_{\mu L} &=&  {i \over M_1} ({g \over 2}E_\mu(x) {\bf 1}- {g'M_1 \over 2m} W^i_\mu(x) \sigma^i )\nonumber \\
A^v_{\mu R} &=& {i \over M_1} g E_\mu(x) ~,~
\varphi= {1 \over M_2}\begin{bmatrix}
h_0 & h^*_1 \cr
-h_1 & h
\end{bmatrix} \nonumber\\
H^\alpha & =& \begin{bmatrix}
h_0 \cr
h_1 \cr
\end{bmatrix},
\end{eqnarray} 
where ${\bf 1}$ and $\sigma^i$ are $2 \times 2$ unit and Pauli matrices with the indices $\alpha, \beta = 1,2$. 

The 5D v-type gauge sector of the 6D Hilbert-Einstein action (\ref{SHE}) now is reduced to the 4D electroweak action as follows
\begin{equation}
S_{ew} = M^2_{Pl} \int dx^4 - {1 \over 4} G^{MK} G^{NL} J^\dagger_{MN} J_KL = \int dx^4 {\cal L}_g(5)
\end{equation}
We have the following reduction of ${\cal L}_g(5)$
\begin{eqnarray}
{\cal L}_g(5) = -{M^2_{Pl} \over 4M^2_1}( g^2F^{\mu\nu} F_{\mu\nu}+ {g'^2M^2_1 \over m^2}G^{\mu \nu} G_{\mu \nu}) + {\cal L}_H &&\\
{\cal L}_(\bar{H}, H) = {\cal L}_{kin}(\bar{H}, H) + V(\bar{H},H)~~~~~~~~~~~~~~&&,
\end{eqnarray}
where
\begin{eqnarray}
&&{\cal L}_{kin}(\bar{H}, H) =  {M^2_{Pl} \over 2M_2^2} ({M^2_{Pl} \over 4 M^2}+ {4 \over M^2_{Pl}} C^\nu C_\nu)\nabla^\mu \bar{H} \nabla_\mu H  \nonumber \\
 &&~~~~~~~~~~~~~~~=~{1 \over 2}(1+ {4 \over M^2_2}C^\mu C_\mu) \nabla^\mu \bar{H} \nabla_\mu H  \\
&& ~~V(\bar{H},H)~~ =  {M^2_{Pl} \over M_2^4}(G^{55})^2 (\bar{H} H - m^2)^2  \nonumber \\
&& ~~=~ {M^2_{Pl}\over M_2^4}[{M^2_{Pl}\over 4 M^2}+ {4 \over M^2_{Pl}} C^\mu C_\mu]^2(\bar{H} H - m^2 M^2_2)^2,\\
&& ~~~~~F_{\mu \nu} = \partial_\mu E_\nu - \partial_\nu E_\mu  \\
&& ~~~~~G_{\mu \nu} =  \partial_\mu W_\nu - \partial_\nu W_\mu + ig_w [W_\mu, W_\nu]   \\
&& ~~~~~\nabla_\mu =  \partial_\mu + {i \over 2}(g A_\mu + g's W^i_\mu \sigma^i) 
\end{eqnarray} 
In order to have the correct factors for the kinetic terms of gauge vectors and Higgs field in the reduced Hilbert-Einstein action, we must have the following relations between the parameters
\begin{eqnarray}\label{pararel1}
g = {M_1 \over M_{Pl}} &~,~& g' = {m \over M_{Pl}}  \nonumber  \\
M^4_{Pl} = 4 M^2 M^2_2 &~,~&  m_H= \sqrt{2} {m M^4_{Pl} \over4 M^2_2 M^2}
\end{eqnarray}
At this point, we have successfully derived all the interaction and Higgs fields as components of the generalized gravity in the noncommutative spacetime ${\cal M}^4 \times Z_2 \times Z_2$. We can choose to couple all interactions and Higgs fields to the chiral spinors in 4D manually as done in Ref.\cite{QuyNhon}. Our model postulate two types $v$ and $w$ of chiral spinors. The $v$-type spinors are the usual quark-lepton fields. We can speculate different scenarios, when the $w$-type spinors couple to the gauge fields. In one scenario, it might might be a candidate of the dark matter, which do not interact with electroweak interactions. 

The Dirac's Lagrangian for spinors ${\bar \psi} (\gamma^\mu \partial_\mu + m) \psi$ is generalized as follows
%\begin{widetext}
\begin{eqnarray}
 {\cal S}_{g-\Psi} =  \int dx^6 \sqrt{-det|g|} {2M \over M_{Pl}} Tr<\Psi|{\cal D}|\Psi> ~~~~&& \cr
~=~ {\cal S}(\Psi^v) + {\cal S}(\Psi^w) + {\cal S}(\Psi^v, \Psi^w)~~~~~~~~~~~~~&& \\
{\cal S}(\Psi^u) = \int dx^5 \sqrt{-det|g|} {2M \over M_{Pl}} (<\Psi^u| \Gamma^A E^M_A(x) (\partial_M ~~&& \cr
+  im_1 A^u_M)|\Psi^u>~~~~~~~~~~~~~~~~~~~~~~~~~~~~ && \\
{\cal S}(\Psi^v,\Psi^w)=  im_1 \int dx^5 \sqrt{-det|g|} {2M \over M_{Pl}}(<\Psi^v|\theta| \Psi^w> \nonumber \\
 - <\Psi^w|\theta| \Psi^v> )~~~~~~~~~~~~~~~~~~~~~~~~~~~~~~~~&&  
\end{eqnarray}
%\end{widetext}

Let us reduce the action ${\cal S}(\Psi^v)$ to the following 4D action
%\begin{widetext}
\begin{eqnarray}
S_5(\Psi^v) &=& \int dx^4 \sqrt{-det|g|}(<\Psi^v| \Gamma^A E^M_A(x)\partial_M|\Psi^v> \cr 
S_g(\Psi^v) &=& \int dx^4 \sqrt{-det|g|} (<\Psi^v| \Gamma^A E^M_A(x) A^v_M)|\Psi^v> \cr
{\cal S}_(\Psi^v) &=& S_5(\Psi^v)+ S_g(\Psi^v).
\end{eqnarray}
%\end{widetext}
In order to obtain the correct Dirac Lagrangian we assume
\begin{equation}
M = {M_{Pl} \over 2},
\end{equation}
which simplifies the relations (\ref{pararel1}) to
\begin{eqnarray}\label{pararel2}
g = {M_1 \over M_{Pl}} &~,~& g' = {m \over M_{Pl}}  \nonumber  \\
M_2 = M_{Pl} &~,~&  m_H= \sqrt{2} m 
\end{eqnarray}
So, the Lagrangian for the Higgs field is reduced to 
\begin{eqnarray}
{\cal L}(\bar{H}, H)&=&{1 \over 2}(1+ {4 \over M^2_{Pl}}C^\mu C_\mu) \nabla^\mu \bar{H} \nabla_\mu H \nonumber \\
&+& (1+ {4 \over M^2_{Pl}} C^\mu C_\mu)^2(\bar{H} H - m^2)^2,
\end{eqnarray}

Let us reduce further the 5D actions $S_5(\Psi^v)$ and $S_g(\Psi^v)$ as follows
\begin{eqnarray}
& & S_5(\Psi^v)  = \int dx^4 \sqrt{-det|g|}(<\Psi^v| \gamma^a e^\mu_a(x)\nabla_\mu|\Psi^v> \nonumber \\ 
&& ~~~~~~~~~~~~~~+ <\Psi^v| \Gamma^5 \Phi|\Psi^v>) \\
&& ~~~~~\nabla_\mu = \partial_\mu + i g_S A^v,  \\
&& S_g(\Psi^v) = \int dx^4 \sqrt{-det|g|} (<\Psi^v_L| \gamma^a e^\mu_a(x) A^v_{\mu L}|\Psi^v_L>  \nonumber \\
&& ~~~~~~~~~~~~~+  <\Psi^v_R| \gamma^a e^\mu_a(x) A^v_{\mu R}|\Psi^v_R>  \nonumber \\
&& ~~~~~~~~~~~~~+ <\Psi^v|\Gamma^a E^5_a(x) A^v_5+ \Gamma^{\dot{5}} E^5_{\dot{5}}|\Psi^v>.
\end{eqnarray}
In order to have to correct factors in the above expressions, one must have
\begin{equation}
g_S = {2m_2 \over M_{Pl}} ~~~~,~~~~ \theta_S = {3\pi \over 8} 
\end{equation}

In a more general form, our unified model has a finite physical field content consisting of four gravity, four Brans-Dicke scalars, eight gauge vectors and two complex Higgs doublets. We have just shown that all the known interactions and Higgs fields can be derived from the generalized Hilbert-Einstein action with specific choices of the physical fields in accordance to our current knowledge and observations in high energy physics. The minimal unified theory has four free parameters: the coupling constants $g, g', g_S$ and the Planck mass $M_{Pl}$. The derived model contains $SU(3)$ color gauge violation term $ (4/M^2_{Pl}) C^\mu C_\mu$ and parity violation angle $\theta_S = 3\pi/8$.

In this unified model, we can speculate that the new $w$-type chiral spinors do not couple to the electroweak interaction and Higgs fields. The w-type matter might still interact with the ordinary gravity and the gluon fields.   

At the moment, we have no reason to assume that the gravitations are different for right and left-handed chiral quark-leptons. However, nothing prevent us from assuming that the gravity is different for the $w$-type matter, which implies the existence of massive gravity, which might have interesting implications in cosmology \cite{DeRham}.

There are still questions in our model to be addressed in the future research. The most important one is what is the energy scale, where the theory becomes valid. In particular, the relations between the coupling constants and the mass parameters must be discussed and explored with more detailed elaborations. The hyperfine structure of Dirac operator must be clarified to describe the mixing parameters and see-saw mechanism for neutrino in a more realistic model.

Our framework has a shortcoming from the mathematical point of view that different treatments of differential forms have been applied in the gauge and gravity sectors. After the manuscript is completed the new calculations have shown that it is possible to use the same wedge product for gravity and gauge theories in NCG, to imply essentially the same results. The additional obtained terms contain only derivatives of the Brans-Dicke scalars, which do not alter the conclusions of this article. The detailed discussions will be given in the subsequent paper \cite{VDD2015}.

Thanks are due to Nguyen Suan Han, Pham Tien Du, Do Van Thanh (Department of Physics, College of Natural Sciences, VNU), Nguyen Van Dat (ITI-VNU) and K.C.Wali (Physics Department, Syracuse University) for their discussion. The supports of ITI-VNU and Department of Physics, College of Natural Sciences, VNU are greatly appreciated.

\end{document}